\documentclass[letterpaper,floatfix,reprint,aps,pre,superscriptaddress]{revtex4-2}
\usepackage{amsmath}
\usepackage{graphicx}% Include figure files
\usepackage{dcolumn}% Align table columns on decimal point
\usepackage{bm}% bold math
\usepackage{color, soul}
\newcounter{MD}
\newcommand*\MD%
{%
\ifnum\theMD>0%
{MD}\fi%
\ifnum\theMD=0%
{Molecular Dynamics (MD) \setcounter{MD}{1}}\fi%
}

\newcounter{SI}
\newcommand*\SI%
{%
\ifnum\theSI>0%
{SI}\fi%
\ifnum\theSI=0%
{Supplementary Information (SI) \setcounter{SI}{1}}\fi%
}

\begin{document}

\preprint{arXiv:yyymmnnn}

\title{Programming Interactions in Magnetic Handshake Materials}

\author{Chrisy Xiyu \surname{Du}}
 \affiliation{School of Engineering and Applied Sciences, Harvard University, Cambridge MA 02139, USA}
\author{Hanyu Alice \surname{Zhang}}
 \affiliation{School of Applied and Engineering Physics, Cornell University, Ithaca, NY, USA}
\author{Tanner \surname{Pearson}}
 \affiliation{Department of Physics, Cornell University, Ithaca, New York, 14853, USA}
\author{Jakin \surname{Ng}}
 \affiliation{Department of Physics, Cornell University, Ithaca, New York, 14853, USA}
\author{Paul \surname{McEuen}}
 \affiliation{Laboratory of Atomic and Solid-State Physics, Cornell University, Ithaca, NY, USA}
\author{Itai \surname{Cohen}}
 \affiliation{Laboratory of Atomic and Solid-State Physics, Cornell University, Ithaca, NY, USA}
\author{Michael P. \surname{Brenner}}
 \affiliation{School of Engineering and Applied Sciences, Harvard University, Cambridge MA 02139, USA}

\date{\today}

%% Nature Materials Letter Guidelines
%%% A Letter reports an important novel research result, but is less substantial than an Article. Letters typically occupy four printed journal pages. This format begins with an introductory paragraph (not abstract) of 150 words maximum, summarizing the background, rationale, main results and implications. This paragraph should be referenced, as in Nature style, and should be considered part of the main text, so that any subsequent introductory material avoids too much repetition of the introductory paragraph. The text is limited to 2000 words, excluding the introductory paragraph, Methods, references and figure legends. As a guideline, Letters allow up to 30 references. Letters should have no more than 3–5 display items (figures and/or tables). A Methods section is published online-only, immediately following the main text and figures. It should be written in such detail that experiments can be reproduced by others.

%%% Letters include received/accepted dates and may be accompanied by supplementary information. Letters are peer reviewed.

\begin{abstract}
The ability to rapidly manufacture building blocks with specific binding interactions is a key aspect of programmable assembly.  Recent developments in DNA nanotechnology and colloidal particle synthesis have significantly advanced our ability to create particle sets with programmable interactions, based on DNA or shape complementarity. The increasing miniaturization underlying magnetic storage offers a new path for engineering programmable components for self assembly, by printing magnetic dipole patterns on substrates using nanotechnology. How to efficiently design dipole patterns for programmable assembly remains an open question as the design space is combinatorially large. Here, we present design rules for programming these magnetic interactions.  By optimizing the structure of the dipole pattern, we demonstrate that the number of independent building blocks scales super linearly with the number of printed domains.  We test these design rules using computational simulations of self assembled blocks, and experimental realizations of the blocks at the mm scale, demonstrating that the designed blocks give high yield assembly.  In addition, our design rules indicate that with current printing technology, micron sized magnetic panels could easily achieve hundreds of different building blocks.
\end{abstract}

\maketitle

\begin{figure}
    \centering
    \includegraphics[width=0.45\textwidth]{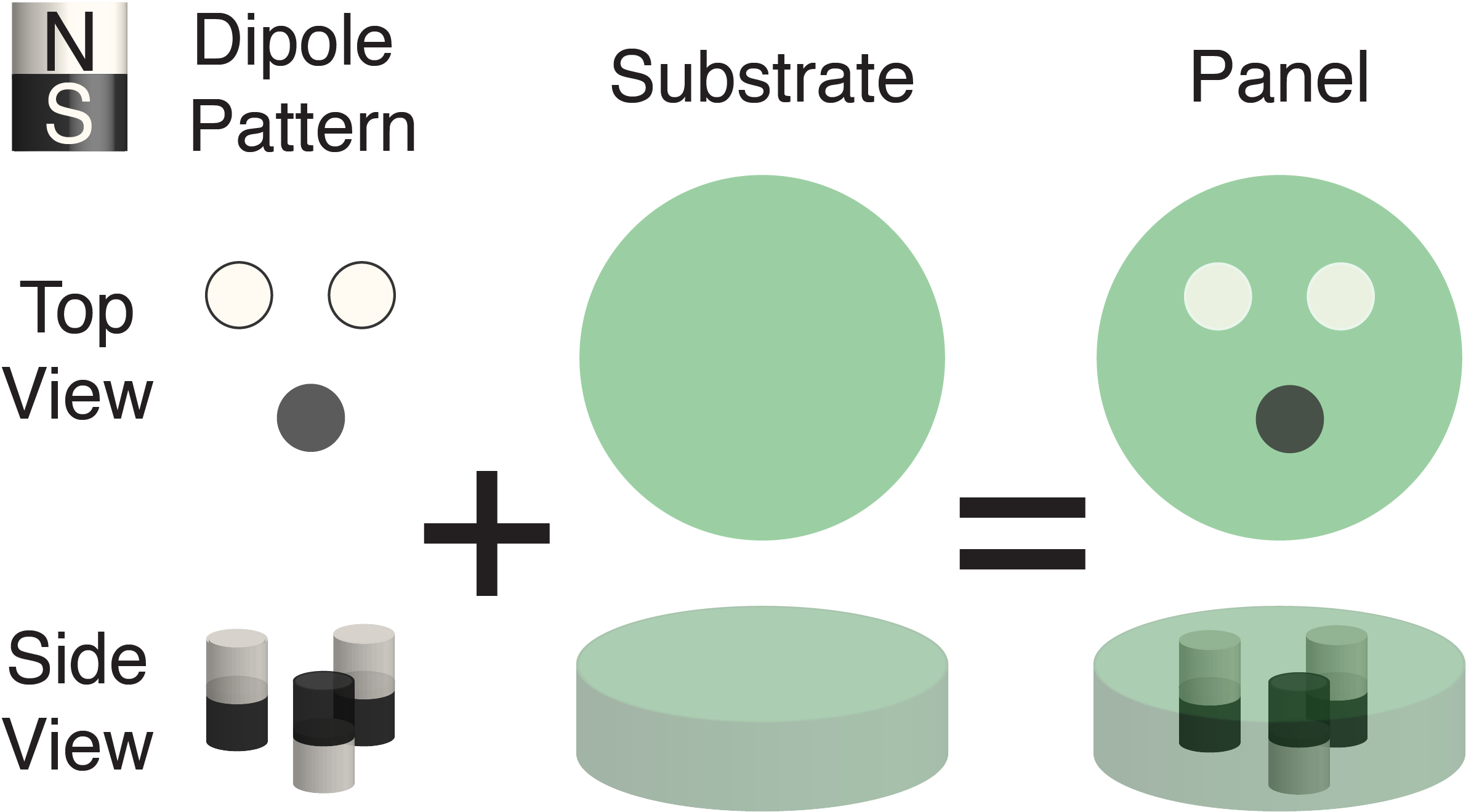}
    \caption{Illustration of general design principle of magnetic handshake panels.  We encode information of specific binding by printing dipole patterns, specifying location and orientation of magnetic dipoles, on substrates.}
    \label{fig:model}
\end{figure}

A key feature of living materials is the inherent programmability of their parts. Complex assemblies require that components with low crosstalk bind to desired partners without binding to others \cite{johnson2011nonspecific}. Such programmability is at the core of biochemical functionality, from protein folding \cite{dobson2003protein, dill2012protein} to self assembly of catalytic protein complexes \cite{kirchhausen2000clathrin, luo2016protein} or multicomponent molecular machines \cite{kay2015rise, erbas2015artificial}.  Biology has managed to create a robust interaction set from a limited set of nucleic and amino acids. Major advances have been made in the engineering of specific interactions that use these biological solutions, by either directly programming nucleic acid assembly \cite{douglas2009self, wagenbauer2017gigadalton} or by using them as specific glues coating nanoparticles \cite{boles2016self, jones2015programmable} or colloids \cite{wang2015crystallization, rogers2016using}. However, these interactions can be hard to program as they are all based on hydrogen bonds, which have a fixed binding energy strength.  An alternative solution for designing programmable building blocks are magnetic handshake materials \cite{niu2019magnetic} whose interactions are governed by magnetic dipole patterns (Fig.~\ref{fig:model}). Such materials can be made over a range of size scales, extending to the nanoscale, where, bootstrapping off Moore's law-like advances in magnetic recording technologies \cite{moser2002magnetic, dorsey2019atomic, cui2019nanomagnetic, hsu2021micromagnetic}, nanotechnology can be used to print magnetic dipoles on substrates. This technology offers an information rich substrate for programming interactions between building blocks to create novel materials, with a potentially large number of programmable building blocks. Dipole patterns can vary in both their strength and their spatial distribution. Over the past decade,  bit sizes have approached 30 nm, the fundamental stability limit of magnetic domains \cite{weller1999thermal, richter1999longitudinal, richter2012thermodynamic}, giving us a combinatorially large space to design interactions.

The major goal of this paper is to elucidate  the basic rules for programming building blocks from magnetic printing. For a given building block design, we need to determine the positions and strengths of the magnetic domains to maximize programmability.  We find these programming rules by combining information theory \cite{huntley2016information} with physical modeling.  Magnetic interactions are simple to describe mathematically and are easy to model, in contrast to electrostatic and chemical interactions. The absence of magnetic monopoles eliminates the complicated screening charge configurations that plague quantitative modeling of electrostatic systems. This simplification makes design iteration computationally efficient and highly predictive.

\textbf{Information Capacity Formalism of Lock-Key Binding Pairs.} The central design requirement for programmability is finding building blocks with low crosstalk. For a system in thermodynamic equilibrium, the yield of an undesired configuration decays exponentially with its energy gap $\Delta$ from the desired configuration \cite{huntley2016information}.  Designing for low crosstalk therefore requires calculating the distribution of binding energies for the components, and optimizing the $\Delta$ distribution to maximize the yield of the desired configuration.

To formulate this optimization problem in terms of the basic building blocks, we consider
 a set of lock and key pairs, denoted as $x_1, x_2, \dots, x_M\in X$ for locks, and $y_1, y_2, \dots, y_M\in Y$ for keys, with the binding energy matrix $E$, where $E_{ij}\equiv E(x_i,y_j)$.  Assuming locks and keys have the same concentrations, the equilibrium  probability of the lock $i$ to bind the key $j$  is $p(x_i,y_j) = e^{-\beta E_{ij}}/Z$, where $\beta$ is the inverse temperature and $Z$ is the partition function.    
A natural metric for quantifying cross talk is mutual information \cite{huntley2016information}, given by
\begin{equation}\label{eq:mutual}
I(X;Y) = \sum_{x_i\in X, y_j\in Y}p(x_i,y_j)\log\frac{p(x_i,y_j)}{p(x_i)p(y_j)}.
\end{equation}
Maximizing the mutual information $I(X;Y)$ over the sets of locks and keys $X, Y$ minimizes the cross talk. The set with the optimal information capacity $I_c=\text{max}(I)$ has effectively $M_\text{c}=\exp({I_c})$  non-crosstalking lock-key pairs.

For a system with a well-defined number of lock and key building blocks and binding energy matrix $E$, we can directly calculate $M_c$ from Eq.~\ref{eq:mutual}.  For experimental systems without  a set number of locks and keys, we can calculate $M_c$ by first computing the energy gap distribution $\Delta$ based on binding energy matrices of random sampled components (see details in \SI).  We define $\Delta$  as $\Delta_{ij}=E_{ij}-s_{ii}$, where $s_{ii}$ is the strength of the on-target binding of lock $i$ and key $i$ and $E_{ij}$ is the off-target binding energy between lock $i$ and key $j$.  Denoting $\rho(\Delta)$ as the distribution of gap energies between on target and off target binding, the maximal number of non-crosstalking pairs ($M_c$)  is given by \cite{huntley2016information}
\begin{equation}\label{eq:mc}
M_c=\frac{(1+\langle e^{-\beta\Delta}\rangle)^2}{\langle\beta\Delta e^{-\beta\Delta}\rangle-\langle e^{-\beta\Delta}\rangle+\langle e^{-\beta\Delta}\rangle^2},
\end{equation}
where $\langle\cdot\rangle$ is the average with respect to $\rho(\Delta)$.
This formula is intuitive:
if we consider the (exponentially weighted) average of the gap $\Delta$ much larger than $k_\text{B} T$ so that $\beta\Delta \gg 1$, then
$M_c$ increases {\sl exponentially} with $\Delta$ as  
\begin{equation}\label{eq:exp}
M_c \sim (\langle\beta\Delta e^{-\beta\Delta}\rangle)^{-1}.
\end{equation}

Using this framework, we can directly formulate an optimization problem of how to encode dipole patterns for a set of magnetic building blocks to maximize programmability.  Here, we measure programmability by evaluating $M_c$ of a set of magnetic patterns, the larger the $M_c$, the better the programmability. Given a potential design for a set of magnetic dipoles on a substrate, we can compute the distribution of gap energies $\rho(\Delta)$, and thus the effective number $M_c$ of programmable building blocks (Eq. \ref{eq:mc}).

\begin{figure}
    \centering
    \includegraphics[width=0.45\textwidth]{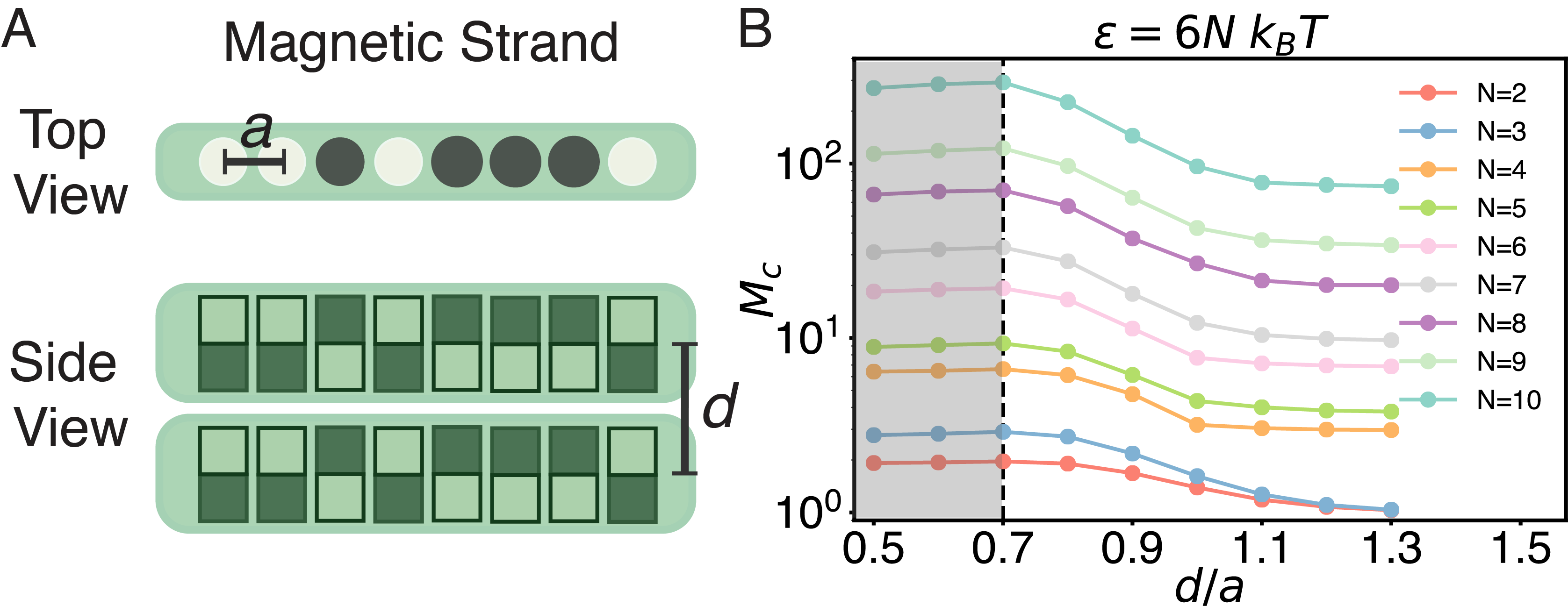}
    \caption{1D Magnetic Strand with Fixed Single Dipole-Dipole Binding Energy. A: sample configuration of a magnetic strand with $N=8$ dipoles.  The distance between any two nearest neighbor dipoles on the same strand is denoted as $a$, while the interaction distance between two different strands is denoted as $d$.  B: the largest effective number ($M_c$) of strands with no crosstalk of magnetic strands of length $N=2-10$ with different $d/a$ ratio.  Beyond $d/a=0.7$, we find that $M_c$ rapidly decreases.  Here, we used $\epsilon=6Nk_\text{B}T$ to show the maximum effect on $M_c$ varying $d/a$ ratio.}
    \label{fig:strand}
\end{figure}

\begin{figure*}
    \centering
    \includegraphics[width=0.9\textwidth]{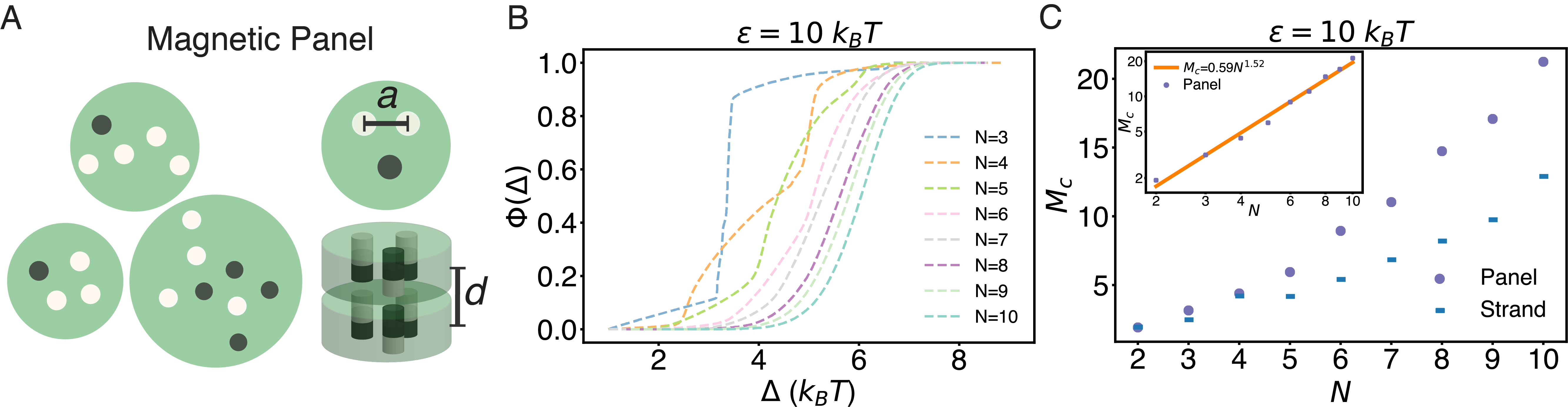}
    \caption{2D Magnetic Panel with Fixed Total Binding Energy. A: three sample dipole patterns with number of dipoles $N = 4, 5, 8$.  The dipoles are placed with the constraint that all nearest neighbor distances for any two dipole pairs must be the same. B: the cumulative distribution functions of $\Delta$ for magnetic panels with $N=3-10$ dipoles with total binding energy $\epsilon=10k_\text{B}T$. C: $M_c$ of magnetic panels and strands with $N$ dipoles for $\epsilon=10k_\text{B}T$.  We see that by allowing the placement of dipoles on a 2D plane, we increased information capacity of our system.  The insert shows the scaling behavior of $M_c$ as a function of $N$ dipoles for panels at $\epsilon=10k_\text{B}T$.  We note that at $\epsilon=10k_\text{B}T$, $M_c$ increases super linearly as a function of $N$.}
    \label{fig:panel}
\end{figure*}

\textbf{Magnetic Interaction Model}. Magnetic interactions depend on the spatial configuration of their dipoles, and in particular the ratio $d/a$, where $a$ is the center to center distance between adjacent dipoles on the same panel, and $d$ is the distance between the centerlines of two panels (Fig.~\ref{fig:strand}A).  By summing over the interactions between every dipole pair, we can write down the binding energy for on-target binding between two complementary magnetic panels.  
\begin{equation}\label{eq:binding}
V=-\frac{\mu_0 m^2}{2\pi d^3}\left[N+k\cdot\frac{d^3}{a^3}\cdot\frac{2d^2/a^2-1}{(1+d^2/a^2)^{5/2}}+\mathcal{F}\left(\frac{d}{a}\right)\right]
\end{equation}
Here $\mu$ is vacuum permeability, $m$ is the magnitude of the magnetic dipole moments, $N$ is the number of dipoles on a single panel, $k$ is the number of second-nearest neighbor aligned dipole pairs minus the anti-aligned dipole pairs, and $\mathcal{F}$ is the (negligible) effect of third or higher neighbor interactions.  Eq.~\ref{eq:binding} implies that the second-nearest neighbor term vanishes when $2d^2/a^2=1$, corresponding to the case where a dipole's magnetic field is perpendicular to its next nearest neighbor (Fig.~\ref{fig:strand}A).  Thus, at $d/a=\sqrt{2}/2$, binding energies between magnetic panels only have contributions from the nearest neighbor interactions and negligible higher order terms.  

\textbf{1D Magnetic Strands}.  To test whether the separation ratio of $d/a=\sqrt{2}/2$ gives rise to a higher information capacity of magnetic panels, we first consider a one dimensional chain of magnetic dipoles, where we can systematically enumerate the entire configuration space of $2^N$ dipole strands with $N$ dipoles. When generating the distinct dipole strand, we remove palindromic sequences corresponding to the same lock-key pair, such as NNNNSSSS and SSSSNNNN.  We enumerate the number of non-palindromic strands for $L\in[2,10]$, giving $[2, 3, 7, 10, 21, 36, 78, 136, 327]$ distinct patterns.  Using Eq.~\ref{eq:mutual}, we directly compute $M_\text{c}$ as a function of $d/a$, by calculating the binding energy matrices of  all possible magnetic strands (Fig. \ref{fig:strand}B). We find that the information capacity $M_\text{c}$ decreases rapidly beyond $d/a\sim 0.7$, confirming our hypothesis that the magnetic strand system has higher information capacity by eliminating second-nearest neighbor interactions.  

Fig.~\ref{fig:strand}B shows that $M_c$ increases exponentially with the number $N$ of dipoles, arising due to the exponential dependence of $M_c$ on the energy gap $\Delta$ (Eq.~\ref{eq:exp}).  Since we computed $M_c$ values by fixing the strength of the individual dipoles, both the on target binding energy and the distribution of energy gap increases linearly with $N$, resulting an exponential increase of $M_c$. This energy dependence highlights a fundamental constraint for programmability.  The timescale of unbinding increases exponentially with interaction strength. Practical design requires the unbinding timescale to be much smaller than the experimental timescale, as otherwise the system will be trapped in non-desired configurations.   Given the timescale of an experiment, we need to choose the binding strengths to be small enough for this condition to be satisfied.  Therefore, the correct optimization problem is to maximize the number of components for a fixed total binding energy.

\begin{figure*}
    \centering
    \includegraphics[width=0.9\textwidth]{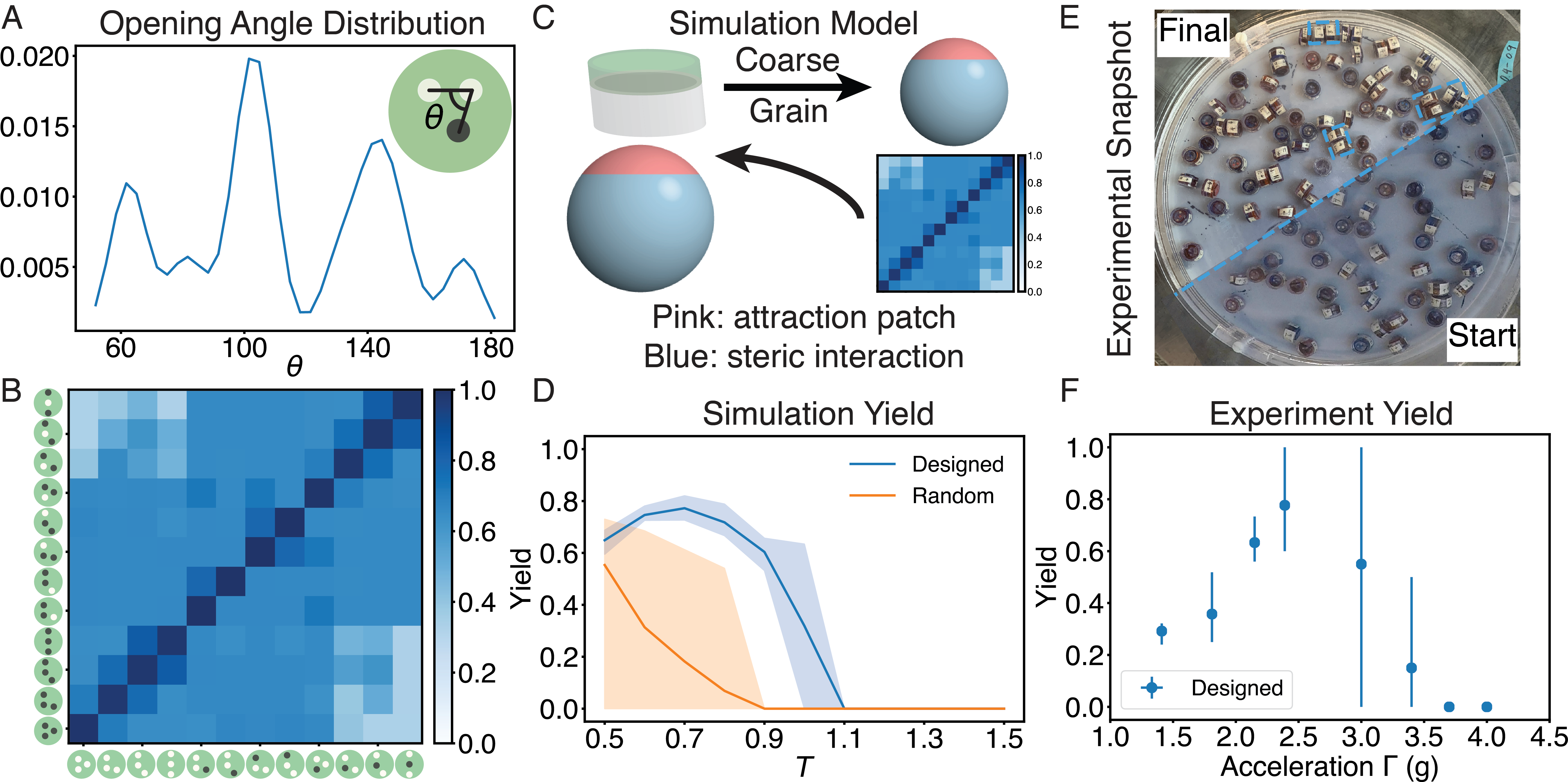}
    \caption{A: the opening angle ($\theta$) distribution for the high information 3-Dipole panels before optimization. B: the binding energy matrix of the designed 3-Dipole Panels.  The binding energies are scaled with the strongest binding energy of the 12 lock-key pairs.  C: the coarse grain process of converting a magnetic panel to a sphere with an attractive patch for simulation.  D: the self-assembly yield of the designed panels compare to the randomly generated ones (not obeying fixed $d/a$ ratio) as a function of simulation temperature. E: start and final state from shaker experiment.  Dimers in blue rectangles form the desired lock-key pairs. F: the self-assembly yield of the designed panels as a function of shaker acceleration (unit in gravitational acceleration $g$).}
    \label{fig:dimer}
\end{figure*}

\textbf{2D Magnetic Panels}. We now turn to the design of dipole configurations on two dimensional panels. We fix the total on-target binding energy $\epsilon$ between any two building blocks to be $10k_\text{B}T$, so a panel with 3 dipoles has individual dipoles with a magnetic moment that is $\sqrt{10/3}$ the dipole moment in a  10-dipole panel.  We consider placements of dipoles on the plane, keeping the constraint $d/a=\sqrt{2}/2$.  We compute $\rho(\Delta)$ by random sampling, over configurations where every dipole has at least one nearest neighbor of distance $a$, with no nearest neighbors smaller than $a$ (Fig.~\ref{fig:panel}A). Using the library, we calculate the equilibrium configuration and total binding energy $E_{ij}$ for each pair of panels in the library. 

Fig.~\ref{fig:panel}B shows the cumulative distribution function for 2D panels with $N=3-10$ dipoles, where we fixed the total binding energy between building blocks to be 10$k_\text{B}T$. The energy gap between on and off target binding increases with increasing $N$. Fig.~\ref{fig:panel}C compares $M_c$ of 1D strands and 2D panels at fixed total binding energy of $\epsilon=10 k_\text{B}T$, where the 2D panels contain much more information.  In addition, the insert of Fig.~\ref{fig:panel}C shows that $M_c \sim N^{1.5}$ for $\epsilon=10k_\text{B}T$, namely the number of effective components increases super linearly with $N$.  This scaling means that even with fixed binding energy between components, we can increase the programmability of our system by printing more magnetic dipoles. This is important, as it suggests that increase in magnetic storage information density can directly result in an increase in information content of panels for assembly without changing the timescale for unbinding.

\textbf{Dimer Self-Assembly}. To validate that the designed dipole patterns can accurately assemble, we carry out both molecular dynamics simulations and experiments comparing the yield of dimer assembly between designed and randomly generated 3-dipole patterns. In the simulations the total on-target binding energy was set to $10k_\text{B}T$.  For the experiments, we used panels with N40 Neodymium magnets separated by $a = 0.28 \text{cm}$ and binding distance $d = 0.2 \text{cm}$, and varied the shaker acceleration to control the effective temperature of the system.  We find the best patterns of 3 dipoles  by iterating through randomly generated patterns, selecting those with high information content.  We then applied symmetry analysis by measuring the opening angle $\theta$ (Fig.~\ref{fig:dimer}A) distribution on all the patterns and identified the optimal symmetries for the best patterns ($\theta$ peaks around 60, 100, 140, and 180 degrees).  Lastly, we generated patterns with said symmetries and iterated over different dipole configurations to get the final designed patterns (Fig.~\ref{fig:dimer}B).  The design protocol yields 12 distinct lock-key pairs (Fig.~\ref{fig:dimer}B) with $M_c=3.46$ at $\epsilon=10k_\text{B}T$,  implying that a theoretical  binding yield of about 68\%.

We instantiate these designed configurations in both simulations and experiments, printing the lock-key pairs onto dimers. We carry out the simulations in HOOMD-blue \cite{anderson2020hoomd, nguyen2011rigid, glaser2020pressure}. We program the interactions between lock-key dimers by putting a small interaction patch on a sphere with the computed binding energy (Fig.~\ref{fig:dimer}C).  The patch is small enough to ensure that only dimers can form in our simulation (see \SI\ for simulation snapshots).  We carry out an ensemble of five different runs of the dimer simulations as a function of temperature, measuring the yield of the desired configurations. We compare this yield curve to that computed from 15 sets of randomly generated panels.  Fig.~\ref{fig:dimer}D shows that the designed panels have significantly higher yield than the random panels, with a maximum yield nearly $80 \%$.
 
For the experiments, we followed a similar protocol. We super glued magnets into laser cut holders and made them into cylindrical panels following the theoretical designs.  We placed these 4 copies of each lock-key pair into a shaker, and measured dimer assembly as a function of shaking amplitude.  The shaking is analogous to the temperature of a thermal bath, providing uncorrelated noise to the individual panels. The number of pairs of each dipole pattern is determined to match the simulation area fraction of 0.2.  To carry out the experiment, we start the shaker at high shaking acceleration to thermalize the panels, and slowly decrease the shaking acceleration to a target amplitude where we count the number of correctly and incorrectly assembled dimers (Fig.~\ref{fig:dimer}E,F). For each shaking amplitude, we carry out five replicas of the experiment.   In Fig.~\ref{fig:dimer}F, we see the yield of dimer assembly from the shaker experiment.  The highest yield is around 80\%, which agrees with simulation (Fig.~\ref{fig:dimer}D).

We note a decrease in yield at lower temperature and shaking acceleration for simulation and experiments respectively.  This arises from the non-equilibrium nature of the dimer self-assembly process at low temperature, where the monomers do not have the chance to freely explore space to find the optimal binding pair, but would bind with anything nearby \cite{williams2019self}.  Our dipole-pattern design strategy assumes equilibrium self-assembly process, so in order to achieve better yield at low temperatures, other strategies can be deployed. 

\begin{figure*}
    \centering
    \includegraphics[width=0.9\textwidth]{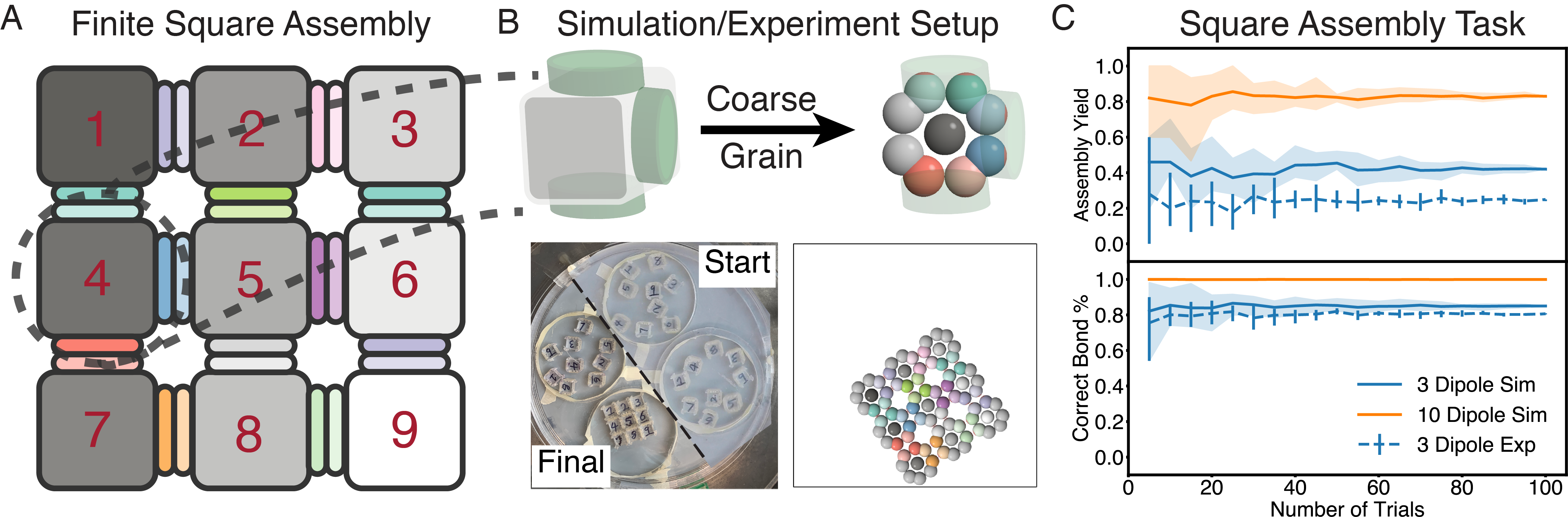}
    \caption{A: model of the 3$\times$3 finite square assembly design.  B: how the fourth building block is represented in an experimental setting and how it is being coarse grained into simulation.  C: the yield of the square assembly using the designed 3-Dipole panels and the 12 chosen 10-Dipole panels.}
    \label{fig:square}
\end{figure*}

\textbf{Finite Square Self-Assembly}. To demonstrate the power of our designed interactions, we show how the magnetic panels can enable finite heterogeneous self-assembly.  We aim to self-assemble a finite 3$\times$3 square (Fig.~\ref{fig:square}A), using 12 different lock key pairs to encode each of the required binding sites.    We instantiate this both using the  12 patterns from the 3-dipole system discussed above, and compare this to a selection of 12 10-dipole patterns (see \SI\ for Panel patterns) generated following the $d/a=\sqrt{2}/2$ design rule.  We selected these 12 10-dipole patterns by identifying the top 12 highest information content patterns from 150 different ones.  For the simulations, we encode the required binding strengths as binding sites on spheres with attractive patches, with each square  building block modeled with a rigid assembly of spheres (Fig.~\ref{fig:square}B).  We first find the optimal temperature for the highest correct binding lock-key pairs by sweeping through a range of different temperatures, and ran 100 independent replica runs at the optimal temperature to measure the yield.  Fig.~\ref{fig:square}C shows the measured percentage of complete square assembly (blue solid line) and correct bonds (yellow solid line) in our simulation.  The correct bond yield is approximately 83\%, consistent with our dimer self-assembly. The yield of perfectly assembled squares is closer to 40\%, with the majority of the errors arising from assemblies with 7 correctly assembled squares, one square having an incorrect bond and one free square.  The 10-dipole panels are significantly more accurate, with the success rate of complete square assembly increasing from 40\% to 80\%, and nearly 100\% rate of correct bonds.

We also carried out experiments for assembling this configuration using panels with three dipole patterns and the shaker. Similar to simulation, we first performed the shaker experiment at different shaking acceleration to identify the optimal amplitude for square assembly (see experimental details in \SI), and we repeated the experiment 100 times at the optimal shaking amplitude. These experiments yielded a correct bond rate of 80\%, whereas 25 out of 100 runs led to perfectly assembled configurations. The correct bond rate is similar to that seen in the dimer experiment and the simulations, while the experimental yield is slightly lower.

\textbf{Discussion}. We demonstrate in this paper a simple design rule for maximizing the mutual information between lock and key pairs.  By fixing $d/a=\sqrt{2}/2$, we maximize the dipole density while minimizing both the cross talk between off-target binding pairs and the distribution width for on-target binding energy over the entire set of dipole patterns.Strikingly, our calculations demonstrate that for two dimensional panels, we increase the programmability by printing more magnetic dipoles, each of which has weaker binding strength. 
Magnetic handshake materials allow us to take full advantage of this design rule, as we can precisely control 
 the strength and the location of the single dipole domains being printed. Moreover, by printing the dipole patterns on two dimensional surfaces -- in contrast to one dimensional chains of e.g. DNA -- we can increase the amount of information that can be encoded in a given building block  (Fig.~\ref{fig:panel}C).  Finally, implementation of our theoretical designs for self-assembly in mm-sized magnetic panels, highlights their applicability to real experimental systems.
 
It is interesting to speculate about the number of potential lock-key pairs that can be created using state of the art nanofabrication technology.
Fig.~\ref{fig:panel}C shows that the number of effective non-crosstalk patterns scale super linearly with the number of dipole domains.  In scaling magnetic handshake materials down to micron-scale panels, it is important to consider the maximum achievable density of non-crosstalking lock-key pairs ($M_c$) given state-of-the-art magnetic recording technology. The current areal bit density of perpendicular magnetic recording media, where one bit represents a magnetic moment oriented into or out of the plane of the disk, is on the order of 1Tb/in$^2$ \cite{marchon2013head}. At this density, one bit comprises an area of 645$\text{nm}^2$, with typical recording layer thicknesses of 10-30nm \cite{hsu2021rotated}. Setting the interaction distance $d$ between two panels as the bit thickness, our $d/a$ design rule dictates a nearest neighbor distance $a$ = 14-32nm. With these constraints, we determine current recording technology could easily achieve a density on the order of hundreds of non-crosstalking lock-key interactions, $M_c$, per 1um$^2$ panels. Thus, magnetic handshake materials offer a new powerful pathway towards the rapid design and creation of programmable interactions for self-assembly of increasingly complex structures.   

\section{Model and Method}
%%Paragraph 1
\subsection{Magnetic Panel Model}
To effectively calculate the binding energy of different designed panel patterns, we approximate all the magnets as magnetic dipoles with a magnetic moment $\vec{m}$, with the direction of the moment  perpendicular to the panel face, either pointing into the plane or out of the plane (see Fig.~\ref{fig:model}).  We assign two geometric parameters to the patterns, $a$ and $d$, where $a$ is the distance between two nearest neighbor dipoles on the same panel, while $d$ is the smallest distance between the center of mass of two interacting panels.  Each panel is characterized by a dimensionless parameter $d/a$  that can be easily translated between theory, simulation and experiment.

%%Paragraph 2
\subsection{Design Strategy}
The magnetic interactions between panels can be calculated directly by summing the well known dipolar interaction energy.  While  nearest neighbor interaction solely depends on $d$, the higher neighbor interactions also depend on  $a$.  We therefore more efficiently explore the design space of possible patterns by generating random dipole pattern with fixed nearest neighbor distance  (see sample patterns in Fig.~\ref{fig:panel}A).   During our design step, we randomly generate 10-150 different panel patterns of $N=3-10$ dipoles, and compute their binding energy matrices for a range of $d/a$ ratios. We repeat this process 10 times to search the parameter space.

\subsection{Simulation and Experimental Setup}
We perform dimer assembly to validate our panel designs and perform a finite square simulation to showcase the possibility in using our magnetic panels for heterogeneous assembly.  For our simulation model, we coarse grain the dipole patterns into an attractive interaction patch \cite{beltran2014phase} with strength based on the binding energy matrix with a WCA potential to mimic steric interactions.  All simulations are performed using HOOMD-blue \cite{anderson2020hoomd, nguyen2011rigid, glaser2020pressure}, while we used Freud \cite{freud2020} for data analysis, and Signac \cite{signac_commat, signac_zenodo} for data management.  More detailed description of the simulation setup can be found in the \SI.

To demonstrate self-assembly with experimental systems, we manufactured panels for dimer and panel assembly. The shaker frequency was set to $25 \pm 1$ Hz for all the experiments while the shaker amplitude was changed with a function generator and amplifier. The dimers are circular disks with diameter 1.27 cm were laser cut from acrylic and glued on top of each other to form a dimer panel with a height of 0.79 cm. For the square assembly experiment, squares with side length 1.52 cm and rounded corners were used. These panels are also cut from acrylic and glued on top of each other to form final panels with height 0.97 cm. A more detailed description can be found in the \SI.

\begin{acknowledgments}
This work was supported by NSF Grants DMR-1921567 and DMR-1921619, Office of Naval Research through ONR N00014-17-1-3029, the Simons Foundation and partially supported by the Cornell Center for Materials Research Grant DMR-1719875.  The computations in this paper were run on the FASRC Cannon cluster supported by the FAS Division of Science Research Computing Group at Harvard University.
\end{acknowledgments}

\textbf{Author Contribution:} C.X.D., HA.Z., T.P., P.L.M., I.C. and M.P.B. designed research; C.X.D.,HA.Z. and J.N. performed research; C.X.D. and M.P.B. contributed new reagents/analytic tools; C.X.D. and HA.Z. analyzed data; and C.X.D., HA.Z., T.P. and M.P.B. wrote the paper.

\bibliography{bibliography}% Produces the bibliography via BibTeX.

\end{document}